# Epitaxial growth of multiferroic Pb(Zr$_{0.57}$Ti$_{0.43}$)O$_3$-Pb(Fe$_{2/3}$W$_{1/3}$)O$_3$ solid-solution thin films and their magnetoelectric effects

D. Lee,[1] Y.-A. Park,[2] S. M. Yang,[1] T. K. Song,[3] Y. Jo,[4] N. Hur,[2] J. H. Jung,[2] and T. W. Noh[1,*]

[1]*ReCOE, Department of Physics and Astronomy, Seoul National University, Seoul 151-747, Republic of Korea*

[2]*Department of Physics, Inha University, Inchon 402-751, Republic of Korea*

[3]*School of Nano and Advanced Materials Engineering, Changwon National University, Changwon 641-773, Republic of Korea*

[4]*Quantum Material Research Team, Korea Basic Science Institute, Daejeon 305-333, Republic of Korea*

We report on epitaxial growth of single-phase [Pb(Zr$_{0.57}$Ti$_{0.43}$)O$_3$]$_{0.8}$[Pb(Fe$_{2/3}$W$_{1/3}$)O$_3$]$_{0.2}$ (PZT-PFW) solid-solution thin films using pulsed laser deposition. X-ray diffraction measurements reveal that the films have a tetragonal structure. The films exhibit ferroelectric properties and weak ferromagnetic responses at room temperature. Magnetoelectric effects were investigated; the nonlinear magnetoelectric coefficient, $β_{33}$, was measured and found to be comparable to those of multiferroic hexagonal manganites, but at least two orders of magnitude smaller than that for polycrystalline PZT-PFW films [A. Kumar *et al.*, J. Phys.: Condens. Matter **21**, 382204 (2009)].

Much research has been focused on solid solutions of conventional and relaxor ferroelectric materials. Ferroelectrics-relaxor ferroelectrics-solid solutions (FRSS) have been used in high-performance piezoelectric actuators and transducer materials (*e.g.*, $PbTiO_3$-$Pb(Zn_{1/3}Nb_{2/3})O_3$ solid solutions).[1,2] Recently, attention has been extended to FRSS materials with magnetic properties. For example, $PbTiO_3$-$Pb(Fe_{2/3}W_{1/3})O_3$ solid-solution thin films have been shown to be both weakly ferromagnetic and ferroelectric, *i.e.*, multiferroic, at room temperature.[3]

Recently, a flurry of studies have been carried out on single-phase multiferroic materials such as $TbMnO_3$ and $BiFeO_3$.[4–11] However, most single-phase multiferroics are difficult to use for practical applications at room temperature, due either to their low Curie temperatures or the small magnetoelectric (ME) effects. More recently, Kumar *et al.*[12] experimentally discovered that $Pb(Zr,Ti)O_3$-$Pb(Fe_{2/3}W_{1/3})O_3$ (PZT-PFW) solid-solution polycrystalline thin films could have very large ME effects at room temperature due to polar nano-regions[13] in relaxor ferroelectrics. They found that the ME coupling of their PZT-PFW films was so large that the remnant polarization could be completely suppressed at room temperature under a magnetic field of 0.5 T. This surprisingly large and promising ME effect was explained by the pseudospin model;[14] however, it is still to be confirmed experimentally.

The multiferroic properties of FRSS is a relatively new research field, and so there have been few studies carried out to date.[3,12,14] Since all of the studies were performed using polycrystalline samples, it has been difficult to measure the physical properties that are dependent on the orientation of the crystal lattice, such as ME tensors, which are essential in order to understand the coupling between charge and spin. In addition, FRSS materials are usually difficult to grow without impurity phases due to the compositional complexity and high volatility of Pb. Therefore, to investigate the multiferroic properties of PZT-PFW, it is highly desirable to grow them as single-phase epitaxial thin films and measure their directionally-dependent physical properties.

Here, we report the epitaxial growth and physical characterization of single-phase $[Pb(Zr_{0.57}Ti_{0.43})O_3]_{0.8}[Pb(Fe_{2/3}W_{1/3})O_3]_{0.2}$ solid-solution thin films. Using the PZT-PFW epitaxial films, which have a well-defined crystalline orientation, we demonstrated a ME effect directly by measuring the quadratic ME coefficient, $\beta_{33}$. Contrary to the reported value of the ME coefficient for polycrystalline PZT-PFW,[12] $\beta_{33}$ in our films was not large enough to suppress the remnant polarization, even with a large magnetic field (up to 4 T).

High-quality PZT-PFW thin films were grown on the $SrTiO_3(100)$ substrates by pulsed laser deposition. We used epitaxially strained $SrRuO_3$ layers as the bottom electrodes for electrical measurements. A PZT-PFW target with excess Pb of 10 at. % was ablated using a 248-nm KrF excimer laser (Lambda Physik) with a laser fluence of 2.5 J/cm$^2$ and repetition rate of 1 Hz. The deposited film thickness was estimated to be ~150 nm. Figure 1(a) shows the X-ray diffraction (XRD) $\theta$–$2\theta$ scans of the PZT-PFW films. Only the (001), (002), and (003) pseudocubic reflections could be seen along with the substrate peaks, indicating the formation of a single-crystalline PZT-PFW phase. We found that the growth window for single-phase PZT-PFW thin films was quite narrow: the growth temperature must be in the range 580–600 °C, the ambient oxygen partial pressure approximately 300 mTorr, and the annealing temperature 450–500 °C. For most other growth conditions, such as higher growth temperatures and/or lower oxygen partial pressures, the films always indicated impurity phases, as shown in the inset of Fig. 1(a).

X-ray crystallography characterization showed that the PZT-PFW thin films were grown epitaxially on the SrTiO$_3$(100) substrates and had a tetragonal crystal structure. As shown in Fig. 1(b), the XRD $\phi$-scan of the PZT-PFW (-103) peak indicated a fourfold symmetry, indicating epitaxial growth of PZT-PFW[100]//SrTiO$_3$[100]. Note that Pb(Zr$_{0.57}$Ti$_{0.43}$)O$_3$ and Pb(Fe$_{2/3}$W$_{1/3}$)O$_3$ have rhombohedral and cubic structures in bulk, respectively.[15,16] If our films had a low-symmetry crystalline structure, such as monoclinic, rhombohedral, or orthorhombic, the X-ray reciprocal space mapping (RSM) data should show a multiple-peak signature,[17] or the peak position should have dependence on in-plane orientations.[18] Figure 1(c) shows the results of RSM around the (-103), (103), (0-13), and (013) peaks of SrTiO$_3$ and PZT-PFW. The RSM data showed only a single peak from the PZT-PFW film, the position of which was independent of the in-plane orientation. Thus, the films had a tetragonal structure. Analysis of the RSM and $\theta$–$2\theta$ data also yielded lattice constants $a$ = 4.021 Å and $c$ = 4.081 Å.

To confirm the relaxor behaviors of our PZT-PFW films, we measured the dielectric permittivity, $\varepsilon_c$, and loss tangent, tan$\delta$, at various temperatures, $T$, and frequencies, $f$. Figures 2(a) and 2(b) show the temperature dependence of $\varepsilon_c$ and tan$\delta$ along the $c$-axis, respectively. The tan$\delta$ data clearly showed two distinct anomalies at ~150 K and ~580 K, denoted by $T_{m,1}$ and $T_{m,2}$, respectively. Note that similar multiple anomalies have been reported in numerous relaxor ferroelectrics,[19–22] including FRSS. Because the films were ferroelectric at room temperature, as shown in Fig. 3(a), the dielectric anomaly near $T_{m,2}$ of ~580 K was attributed to a paraelectric-to-ferroelectric phase transition. On the other hand, the dielectric anomaly near $T_{m,1}$ (at ~150 K) corresponded to typical relaxor behaviors. The peaks in $\varepsilon_c$ and tan$\delta$ moved to higher temperatures and became broadened with increasing frequency.

The frequency dependence of the dielectric anomalies can be explained in terms of relaxor ferroelectrics.[13] According to the standard relaxation theory,[19,23] the dielectric relaxation time, $\tau$, near the anomaly temperature is given by $\tau = 1/2\pi f$. From the measurements of tan$\delta$ shown in Fig. 2(b), we evaluated $\tau(T_{m,1})$. For relaxor ferroelectrics, the temperature dependence of $\tau$ is known to follow the Vogel–Fulcher (VF) relationship,[24] *i.e.*, $\tau = \tau_0 \exp[E_a/k_B(T_{m,1} - T_0)]$, where $\tau_0$, $E_a$, and $T_0$ are the characteristic relaxation time, activation energy, and VF freezing temperature, respectively. As shown in Fig. 2(c), the experimental values of $\tau(T_{m,1})$ were described well by the VF relationship with $T_0 = 135 \pm 5$ K. The values of $\tau_0$ and $E_a$ were approximately $4\times10^{-10}$ s and 0.05 eV, respectively, which were comparable to those of other relaxor ferroelectric materials.[3,25,26]

Our epitaxial PZT-PFW films were found to be ferroelectric. Figure 3(a) shows ferroelectric polarization versus electric field (*P–E*) hysteresis loops measured using a *T-F* analyzer (aixACCT) at 2 kHz. At room temperature, the remnant polarization and coercive field were ~50 μC/cm$^2$ and ~0.14 MV/cm, respectively. These values are about twice those for polycrystalline PZT-PFW films.[12] We also investigated the piezoresponse hysteresis loop by using a piezoelectric force microscope at room temperature. Figure 3(b) indicates that the piezoresponse could be clearly switched.

To investigate the magnetic properties of our PZT-PFW films, we measured the magnetization, *M*, using superconducting quantum interference devices (Quantum Design,

MPMS). The PZT-PFW epitaxial thin films showed a weak ferromagnetic response at room temperature (not shown here), as has been found previously.[3,25,26] Considering the inherent antiferromagnetic interaction among $Fe^{3+}$-O-$Fe^{3+}$ spins in PFW,[12] the observed weak ferromagnetic response might come from the canted $Fe^{3+}$ spins[3] or cluster magnetization.[12,25,26] Although the origin of the weak ferromagnetic response remains to be elucidated, the existence of a weak ferromagnetic response confirms that our PZT-PFW epitaxial thin films were multiferroic.

To investigate the ME effects of our epitaxial PZT-PFW films, we measured the magnetocapacitance ($\Delta\varepsilon_c(H_z)/\varepsilon_c(0) = \varepsilon_c(H_z)/\varepsilon_c(0) - 1$), *i.e.*, the change of capacitance under a magnetic field. $\Delta\varepsilon_c(H_z)/\varepsilon_c(0)$ was measured by varying the magnetic field along the *c*-axis from −6 to 6 T. Figures 4(a)–4(c) show the magnetocapacitance of our films at different temperatures. At room temperature, the magnetocapacitance showed almost quadratic behavior, consistent with the reported behavior of polycrystalline PZT-PFW films.[12,14] However, when $T < 150$ K, the magnetocapacitance appeared almost linear. It is also useful to compare the magnitude of our magnetocapacitance values with those of other multiferroics. $\Delta\varepsilon_c(H_z = 6.0 \text{ T})/\varepsilon_c(0)$ was about 0.06 % at 100 K and 0.02 % at 300 K. Note that these values are comparable to those of multiferroic hexagonal manganites.[8] However, they are much smaller than the reported magnetocapacitance (almost 100 %) of polycrystalline PZT-PFW films.[12]

In the Landau formalism,[12] the quadratic ME coupling can be expressed by the free energy term $F = -(\lambda_{ij}/2)P_i^2 M_j^2$. From this term, the magnetocapacitance for the $H_z$-field can be approximated as[27,28]

$$\frac{\Delta\varepsilon_c(H)}{\varepsilon_c(0)} = \left(\frac{\partial^2 F}{\partial E_z^2}\bigg|_{E_z=0}\right)/\varepsilon_c = -\lambda_{33}\varepsilon_0\chi_{e,3}M_3^2 \cong -\lambda_{33}\varepsilon_0\chi_{e,3}\chi_{m,3}^2 H_z^2 = \beta_{33}H_z^2$$

, (1)

where $\chi_{e,3}$ and $\chi_{m,3}$ are the dielectric and magnetic susceptibilities along the *c*-axis, respectively. Note that Eq. (1) is valid when $\chi_{m,3} \ll 1$. As shown in Fig. 4(d), $\chi_{m,3}$ remained small for $T > 150$ K, which is why we could observe the quadratic behaviors in this temperature region. Using the magnetocapacitance data, we estimated that $\beta_{33}$ was $6.5\times10^{-6}$ T$^{-2}$ at 300 K and $1.9\times10^{-5}$ T$^{-2}$ at 250 K. Figure 4(d) shows that $\chi_{m,3}$ became large for $T < 150$ K. In this case, $M_3$ was no longer linearly proportional to $H_z$, and so the magnetocapacitance should deviate from the quadratic behavior observed at higher temperatures.[27,28]

It is important to check the magnitude of the magnetic field required for our epitaxial PZT-PFW films to experience strong suppression of the remnant polarization. Note that for polycrystalline PZT-PFW films, a magnetic field of only about 0.5 T was required.[12,14] According to the pseudospin model,[14] the critical magnetic field, $H_c$, required for the polarization suppression can be estimated from

$$H_c^2 = \frac{T-T_0}{T_0}\frac{1}{|\langle\lambda_{ij}\varepsilon_0\chi_{e,i}\chi_{m,j}^2\rangle|} = \frac{T-T_0}{T_0}\frac{1}{|\langle\beta_{ij}\rangle|}$$

, (2)

where $\langle\beta_{ij}\rangle$ is an average of components of the quadratic ME tensor $\beta_{ij}$. With the experimental value of $\beta_{33}$ for our epitaxial films, we estimated that $H_c$ should be as large as 20 T, even very close to $T_0$, i.e., $(T - T_0)/T_0 = 0.01$. Due to the orders of magnitude smaller value of $\beta_{33}$, the estimated $H_c$ was much higher than that of polycrystalline films on Pt/TiO$_2$/SiO$_2$/Si substrates (~1 T),[12] and indicates that the remnant polarization of our epitaxial PZT-PFW film could not be suppressed by applying a magnetic field of any practical magnitude along the *c*-axis. As shown in Fig. 5, we were not able to observe any significant suppression of the remnant polarization in our PZT-PFW epitaxial films under magnetic fields of up to 4 T.

The origin of the large difference in the magnitude of the ME effects between our epitaxial and reported polycrystalline films[12] is not clear. However, we can put forward two possible explanations. First, $β_{33}$ might be much smaller than those of other components of $β_{ij}$ with $(i, j) ≠ (3, 3)$ in the quadratic ME tensor. This could induce a large difference in magnitude between $β_{33}$ and $<β_{ij}>$. A second possibility is the clamping effect from the substrate. Substrate-imposed mechanical clamping is known to suppress both the piezoelectric response and magnetoelectric coupling mediated by lattice deformation in thin-film-on-substrate geometries.[29,30] In particular, quadratic ME coefficients $β_{ij}$ also might be reduced by the substrate-imposed clamping, since they could be proportional to electro- and magnetostrictive properties.[12,14] Further investigation to tailor the ME properties of FRSS by varying the strain and/or composition is highly desirable.

In summary, we fabricated single-phase tetragonal PZT-PFW epitaxial thin films by pulsed laser deposition. Epitaxially grown PZT-PFW thin films showed a clear signature of room-temperature multiferroicity. We were able to observe quadratic magnetocapacitance effects and measure an associated coupling coefficient. The magnetoelectric effects of our epitaxial films were much smaller than those reported earlier in polycrystalline PZT-PFW films.

This work was supported by Korea Science and Engineering Foundation (KOSEF)

grant 2009-0080567 funded by the Korean Ministry of Education, Science, and Technology (MEST). D.L. acknowledges support from the TJ Park Doctoral Fellowship and Seoul City.

[27]We roughly assume $M \propto H$, which is true for small values of $H$ and $M$. However, for (weak) ferromagnets with a sufficiently large $M$, the Weiss model gives $M \propto H^{1/3}$ (Ref. 28). Thus, the abrupt increase of $\chi_m$ below ~150 K (Fig. 4(d)) might explain the nearly linear $H$-dependence of $\Delta\varepsilon(H)$ at low temperatures (Fig. 4(a)), since $\Delta\varepsilon(H) \propto M^2 \propto H^{2/3}$.

<FIGURE CAPTIONS>

FIG. 1. (Color online) (a) XRD $\theta$–$2\theta$ scans of the PZT-PFW films on SrRuO$_3$/SrTiO$_3$ substrates. The inset in (a) shows a $\theta$–$2\theta$ scan of films deposited at higher temperature (650 °C) and lower $P_{O2}$ (100 mTorr). (b) $\phi$-scan of the (-103) Bragg peaks for PZT-PFW films. (c) RSM around the (-103), (103), (0-13), and (013) Bragg reflections from the SrTiO$_3$ substrate and PZT-PFW films.

FIG. 2. (Color online) (a) Temperature dependence of $\varepsilon_c$ measured at frequencies of $10^2$, $10^3$, $10^4$, $10^5$, and $10^6$ Hz. (b) Temperature dependence of tan$\delta$. (c) The Vogel–Fulcher relationship between $\tau$ and $T_{m,1}$; the red dashed line is a linear fit from the Vogel-Fulcher relationship.

FIG. 3. (Color online) (a) $P$–$E$ loops measured at 70, 130, 190, and 300 K. (b) Piezoresponse hysteresis loop measured at 300 K.

FIG. 4. (Color online) (a)–(c) Magnetocapacitance data of the PZT-PFW films measured at 1 MHz for temperatures of 100, 250, and 300 K, respectively. (d) Temperature dependence of the inverse magnetic susceptibility (1/$\chi_m$).

FIG. 5. $P$–$E$ loops measured at 150 K, as functions of the applied magnetic field.

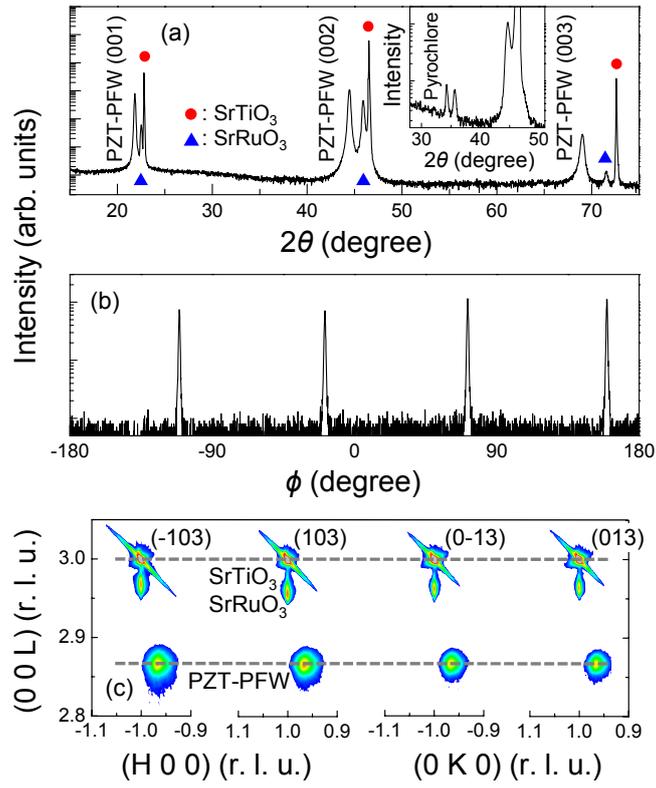

FIG. 1. D. Lee *et al.*

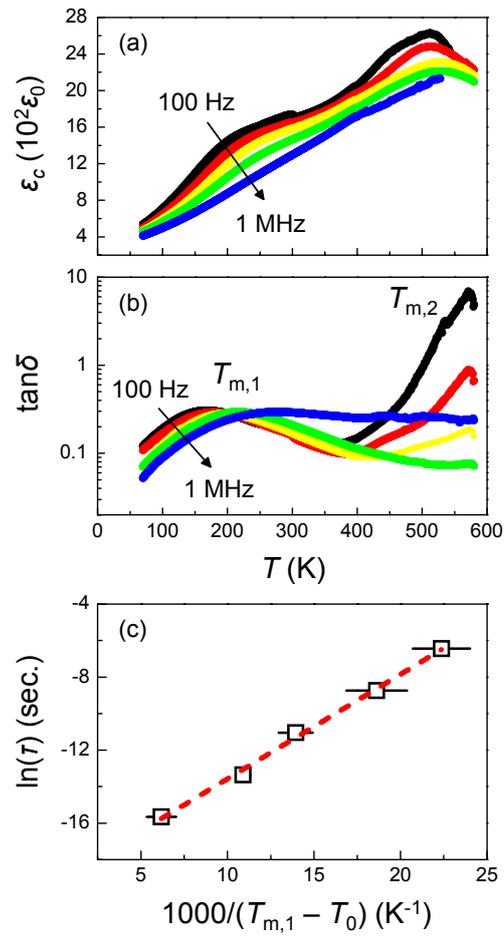

FIG. 2. D. Lee *et al.*

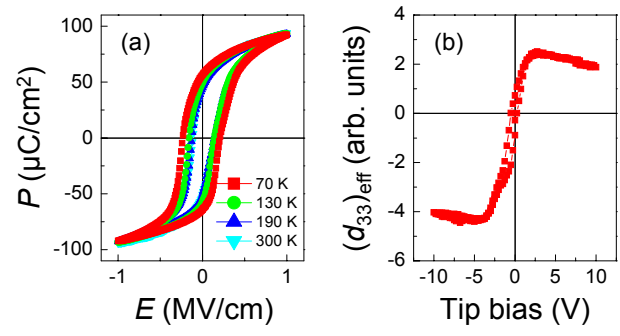

FIG. 3. D. Lee *et al.*

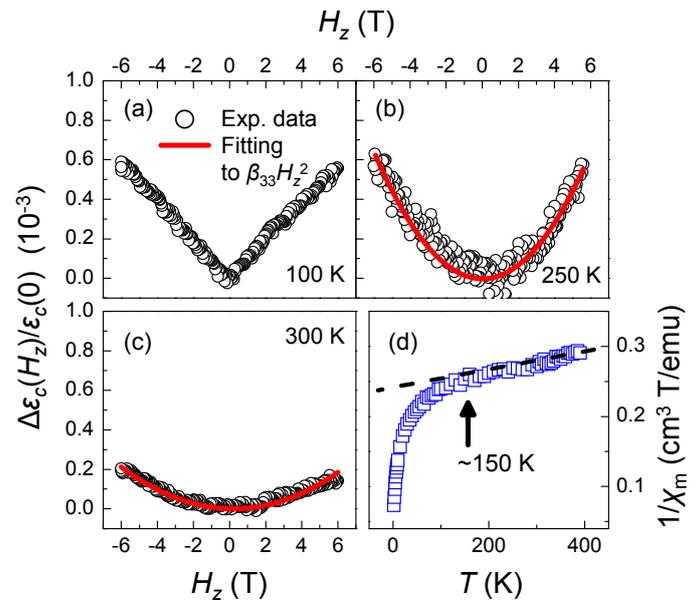

FIG. 4. D. Lee *et al.*

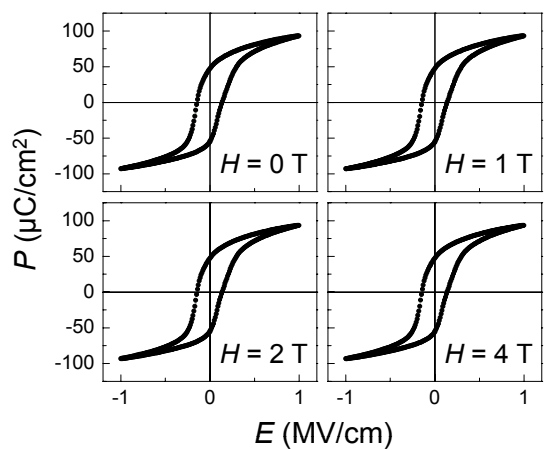

FIG. 5. D. Lee *et al.*